# AI Assisted Method for Efficiently Generating Breast Ultrasound Screening Reports

Shuang Ge [1], Qiongyu Ye [2], Wenquan Xie [1], Desheng Sun [3], Huabin Zhang[4], Xiaobo Zhou[5], Kehong Yuan [1]

1 Graduate School at Shenzhen, Tsinghua University, Shenzhen, China
2 Shenzhen City Baoan District Women's and Children's Hospital, Shenzhen, China.
3 Shenzhen Hospital of Peking University, Shenzhen, China.
4 Beijing Tsinghua Changgung Hospital, Tsinghua University, Beijing, China
5 School of Biomedical Informatics, University of Texas Health Sciences Center at Houston, Houston TX 77030, USA
Corresponding author: Kehong Yuan (yuankh@sz.tsinghua.edu.cn) ; Huabin Zhang (Huabinzhang@bjmu.edu.cn)

*Address reprint requests to: Kehong Yuan, Tsinghua Shenzhen International Graduate School, University Town of Shenzhen, Nanshan District, Shenzhen 518055 P.R. China

## Abstract

**Background**: Ultrasound is one of the preferred choices for early screening of dense breast cancer. Clinically, doctors have to manually write the screening report which is time-consuming and laborious, and it is easy to miss and miswrite.

Aim: We proposed a new pipeline to automatically generate AI breast ultrasound screening reports based on ultrasound images, aiming to assist doctors in improving the efficiency of clinical screening and reducing repetitive report writing.

**Methods**: AI was used to efficiently generate personalized breast ultrasound screening preliminary reports, especially for benign and normal cases which account for the majority. Based on the preliminary AI report, doctors then make simple adjustments or corrections to quickly generate the final report. The approach has been trained and tested using a database of 4809 breast tumor instances.

**Results**: Experimental results indicate that this pipeline improves doctors' work efficiency by up to 90%, which greatly reduces repetitive work.

**Conclusion**: Personalized report generation is more widely recognized by doctors in clinical practice compared with non-intelligent reports based on fixed templates or containing options to fill in the blanks.

## Ⅰ Introduction

According to the 2020 Global Cancer Data(GLOBOCAN) statistical report, the proportion of breast cancer of new cases in 185 countries around the world is 11.7%, which has surpassed lung cancer (11.4%) as the most common cancer. The incidences of breast cancer in developed and developing countries are respectively 55.9/100,000 and 29.7/100,000[1]. The World Health Organization (WHO) has defined early breast cancer as a curable disease. Practice has shown that early screening for breast cancer can significantly reduce mortality[2-5]. The main methods of breast cancer screening are ultrasound, mammography, and MRI. Ultrasound can find more aggressive and smaller lesion and has unique advantages in the examination of dense breast tissue. It has become the preferred breast cancer screening method for dense breast with its noninvasive, convenient and low-cost characteristics[6, 7].

The steps of breast ultrasound screening usually are as follows: the sonographer performs a routine examination and saves the suspicious images. Then the doctor will diagnose according to the saved images and have to give some descriptive sentences as a report. However, current ultrasound screening reports are usually unstructured, and a few structured templates are also with the default description. As a result, doctors need 5 to 10 minutes averagely to complete an ultrasound report, which takes up a lot of time[8]. To solve this problem, Ortega-Loubon et al.[9] explained the importance of structured case reports. Wei M et al.[10] extracted the characteristic information of the lesion from the report to generate structured data. Although the incidence rate of breast cancer is relatively high, in clinical screening, only 0.05% of cases are "malignant", which means the screening results of "benign" and "normal" account for the vast majority[3]. Therefore, research focusing on benign and normal breast ultrasound AI reports is of great significance in terms of optimizing clinical workflow .

In recent years, research on the automatic generation of AI-assisted diagnostic reports has been increasing, but most of the researches have focused on standardized medical images such as MRI, CT, and X-rays images. Although many related studies on the classification of benign and malignant ultrasound images have been proposed[11-15], there are no outstanding results for the automatic generation of ultrasound screening reports. In the research of breast MRI report assistant system, He Y J et al.[16] used Fuzzy C-means (FCM) to locate and classify the breast area, and compute the tumor shape and margin characteristics according to the segmentation results of the floating support-pixel correlation statistical method(FSCSM). Finally, these parameters were used to assist doctors in generating diagnostic reports. Syeda-Mahmood et al.[17] used the Fine Finding Label (FFL) to compare and extract the information in the clinical report library of chest X-rays to automatically generate diagnostic reports. Gale, W et al.[18] extracted key information from medical images, converted the key information into text descriptions, and finally automatically generated an interpretable diagnostic report for pelvic X-rays.

In this article, based on the BI-RADS ultrasonographic descriptors and structured mode, we proposed AI to efficiently generate breast ultrasound screening reports according to ultrasound images, especially for benign and normal cases. Then the doctor can quickly generate the final report by making simple adjustments or corrections. This method aims to reduce the repetitive workload of doctors, shorten the time for generating diagnostic reports, and improve doctors' work efficiency.

## Ⅱ Materials and Methods

### A. Methods Overview

The model proposed in this paper to assist doctors in efficiently generating an ultrasonic breast screening report is shown in Fig. 1. The model combines multiple classifiers. For each image, feature terms of benign tumor and a diagnostic reference to classify the tumor as benign or malignant are generated and presented in the preliminary report. The doctor can further determinate the tumor as benign or malignant. If the tumors are benign, the doctor can then briefly scan or slightly change according to the results of the preliminary AI report to output the final screening report. For tumors judged by the doctor as malignant, the doctor needs to type descriptions to generate the manual report. If the images are identified normal, the doctor can directly give the conclusion of "normal" without additional feature description.

The preliminary AI results reported in AI-assisted model mainly consists of the following three sections : (1) The discrimination of benign or malignant tissue, (2) structured descriptions based on the shape, internal echo pattern and posterior acoustic features of the lesion; (3)fixed feature description lexicon including boundary, orientation and margin, etc.

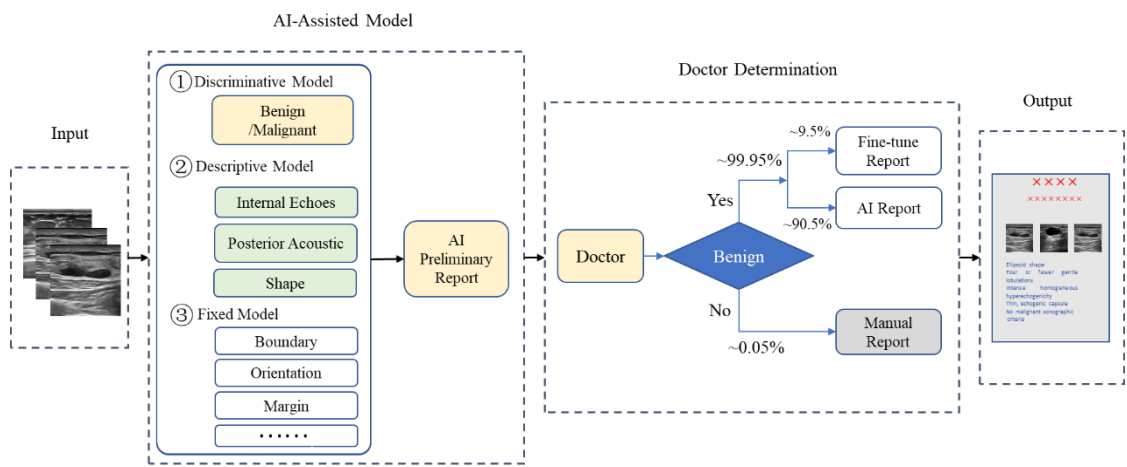

Figure 1. System Architecture Diagram. The preliminary AI report in AI-assisted model mainly consists of the following three sections :(1) The discriminant of benign or malignant tissue; (2) The structured descriptions based on the shape, internal echo pattern and posterior acoustic features of the lesion; (3)The fixed feature description lexicon including boundary, orientation and margin, etc. Studies show that 99.95% cases of breast cancer screening is benign or normal, and only 0.05% suffer from breast cancer. Therefore, it is meaningful to generate AI preliminary reports for the majority.

## 1. Discriminative Model: Benign or Malignant

A deep network structure with automatic feature extraction was used to train the classifier. In the training phase, images labeled as benign or malignant were trained end-to-end. Due to the small scale of medical datasets and the lack of sufficient information to establish relationships between low-level semantic features, the performance of a pure transformer network based on ViT and DeiT is not satisfactory in medical image classification[19, 20]. Therefore, we apply transformer and CNN(Convolutional Neural Networks), and has achieved good results in the classification of benign and malignant breast tumor, indicating that the combination of transformer and CNN is expected to assist in more practical problems in the field of medical images analysis. Additionally, the pre-trained weights on Imagenet1k were used to fine-tune the model for further improvement.

The model based on the combination of CNN and transformer is shown in Fig. 2. It is mainly composed of the following parts: (1)CNN extracts low-level visual and semantic features. Residual connection can effectively prevent gradient from disappearing , which is helpful to improve the effectiveness of feature extraction[21]. (2)In the transition part from CNN to transformer, the linear projection layer of image patches is used to represent the image information in the form of a vector, at the same time, position embedding is added to the patch embedding, and a learnable category embedding vector with the same dimension is concatenated with the embedding of input images.(3)The attention channel is adopted to extract visual information between image patches in transformer module, and the self-attention mechanism gives higher weight to important information to strengthen local attention between patches.(4)Layer normalization and multi-layer perception are used to encode the features, and finally the benign and malignant normalized probability scores are output through the softmax layer.

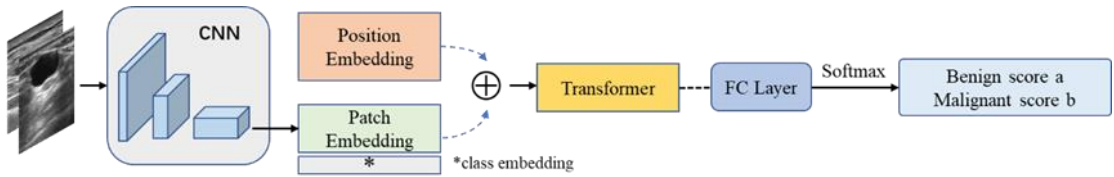

Figure 2. transformer + CNN Network structure diagram. Our system was the first one to apply transformer in ultrasound images analysis and it performed quite well in classifying benign and malignant cases.

## 2. Discriminative model based on focal features

Although we have obtained the results of the benign and malignant judgment of AI-assisted diagnosis in Section A.1., benign and normal cases account for the majority of the actual screening work, which makes it difficult to reduce the workload of doctors with the model of pure benign or malignant judgment. Therefore, based on the key features of ultrasound images, this paper constructed a structured report generation model.

Compared with the errors caused by subclassification, which in turn increase the workload of doctors, we implemented word frequency statistics according to the features of the reports and selected the top-10 features as the labels, while ignoring other special cases. Finally, the terms we selected are shown in Table 1 according to American Society of Radiology BI-RADS ultrasonographic descriptors[22] and clinical experience.

Table 1 Key terms in benign reports

| Shape | Internal Echoes | Posterior Acoustic Feature | Boundary | Orientation | Margin |
|---|---|---|---|---|---|
| Oval/Round | Homogeneous | Enhancement | Abrupt | Parallel | Circumscribed |
| Irregular | Anechoic | No posterior features | | | |

For every lesion, the model generates structured benign descriptors based on the shape, the internal echoes, and the posterior features of the tumor.

In this paper, the CNN method was adopted for feature extraction, and Resnet50 was used as the backbone of the network. CNN can be regarded as a powerful feature extractor, learning end-to-end mapping function from input image to output features. The feature vectors from the full-connection layer were input into the SVM(Supported Vector Machine)[23] classifier for training.

The basic model we used here is a linear classifier defined in the eigenspace with the largest interval, its learning strategy is to maximize the discrepancy interval, which can be transformed into a convex quadratic programming problem. At the same time, the VGG(Visual Geometry Group)[24] method was used for end-to-end training due to the ability of deep feature extraction. VGG16 model contains 13 convolutional layers and 3 fully connected layers, we also used transfer learning algorithm to fine-tune parameters in our own datasets

### *3. Fixed Model*

As can be seen in Table 1, benign cases have the same characteristics in three aspects, including the boundary, orientation and the margin of the tumor, while there are only major differences in the shape, internal echoes and posterior acoustic features. After generating structured benign descriptors in Section A.2., other features (including boundary, orientation, and margin) are given by default to construct the preliminary AI report.

### *B. Datasets*

A total of 4128 breast ultrasound images were collected, including 2064 benign images and 2064 malignant images(Table 2). The images were scanned from different manufacturers, including Philips, Mindray, Hitachi, GE, etc., which increases the diversity of the data. Five sophisticated doctors labeled the data independently, each of whom also examined the images labeled by others. Among them, 873 ultrasound reports with benign diagnosis were selected for structural information extraction. Images without description or with special description were excluded from the experiment. After reappraisal by doctors, the images with exact description were selected according to the characteristics of lesions.

Table 2 Dataset of all features

| Feature | Training set | | Validation set | | Total |
|---|---|---|---|---|---|
| Posterior Acoustic Feature | Enhancement (class 0) | No posterior features (class 1) | Enhancement | No posterior features | |
| | 57 | 516 | 15 | 130 | 718 |
| Internal Echoes | Homogeneous (class 0) | Anechoic (class 1) | Homogeneous | Anechoic | |
| | 296 | 224 | 74 | 56 | 650 |
| Shape | Irregular (class 0) | Oval/Round (class 1) | Irregular | Oval/Round | |
| | 30 | 511 | 8 | 127 | 676 |
| Benign/ Malignant | Benign (class0) | Malignant (class1) | Benign (class0) | Malignant (class1) | |
| | 1869 | 1869 | 195 | 195 | 4128 |

In addition, the benign reports of other five clinical doctors were selected to further test the results of the trained model, and the images involved in the test set were not included in the training set or the validation set, to verify the generalization ability of the model. The statistics of the ultrasound images of each doctor are shown in Table 3, and the inadequately described samples have been discarded.

Table 3 Number of samples of each doctor

| NAME | Internal Echoes | Posterior Acoustic | Shape | Boundary | Orientation | Margin | Total |
|---|---|---|---|---|---|---|---|
| Doctor1 | 45 | 43 | 46 | 46 | 46 | 46 | 272 |
| Doctor2 | 41 | 41 | 41 | 41 | 41 | 41 | 246 |
| Doctor3 | 45 | 42 | 45 | 45 | 45 | 45 | 267 |
| Doctor4 | 32 | 32 | 32 | 32 | 32 | 32 | 192 |
| Doctor5 | 26 | 26 | 26 | 26 | 26 | 26 | 156 |
| **Total** | 189 | 184 | 190 | 190 | 190 | 190 | 1133 |

## *C. Parameters setting*

The discriminative model was optimized by the adaptive gradient descent method and the cross-entropy loss function was used as the objective function. In order to reduce overfitting, the weight decay coefficient was set to 0.0001. Parameters from pre-trained models were also re-used in this experiment.

To solve the problem of unbalanced classified samples, data augmentation methods such as rotation, brightness adjustment, and contrast adjustment were used. Table 4 shows the whole dataset after augmenting. Features were extracted by CNN through end-to-end training. The extracted features were then fed into SVM classifier for further training. At the same time, we also adopted VGG19 network to extract and classify the features and obtain a better classification result.

Table 4 Dataset of augmentation

| Feature | Internal Echoes | Posterior Acoustic Feature | Shape | Total |
|---|---|---|---|---|
| Train | 1256 | 1180 | 966 | |
| Validation | 314 | 288 | 242 | |
| Test | 189 | 184 | 190 | |
| Total | 1759 | 1652 | 1398 | 4809 |

## *D. Metrics*

We use standard Precision(P), Recall(R), and F1 score ($\beta = 0.9$) to evaluation the performance of our classification model. Receiver operating characteristic (ROC) curves and the area under the curve (AUC) were also discussed below. Moreover, confusion matrix was computed, among which $TP, FP, FN, TN$ represent true positive, false positive, false negative and true negative respectively.

The accuracy was used to evaluate the classification results of the model for each feature, which is calculated as follows:

$$accuracy = \frac{TP+TN}{TP+FP+FN+TN}$$

Furthermore, it was also used as a quantitative index for the improvement of the work efficiency. Due to the inaccuracy of the model, the doctor needs to click the mouse for manual input or select other features through options. Most of the time, if the model is exactly correct, the doctor does not need to make any changes, which greatly improves the efficiency of report writing.

# Ⅳ Result

## A. Discriminative Model: classification of "benign" and "malignant"

Compared with transformer baseline model and Resnet50 baseline model, transformer + CNN achieved almost the same performance as the Resnet50 network with or without transfer learning. Results of different methods are listed below(Table 5).

Table 5 results of different methods

| Model | P | R | F1 | AUC |
|---|---|---|---|---|
| transformer | 0.8643 | 0.8821 | 0.8722 | 0.9391 |
| transformer pretrained | 0.8325 | 0.8418 | 0.8363 | 0.9260 |
| Resnet50 | 0.8821 | 0.8821 | 0.8821 | 0.9522 |
| Resnet50 pretrained | **0.9581** | **0.9385** | **0.9492** | **0.9799** |
| transformer +CNN | 0.8895 | 0.8667 | 0.8791 | 0.9556 |
| transformer + CNN pretrained | **0.9574** | **0.9231** | **0.9418** | **0.9755** |

Fig. 3. draws the classification confusion matrix of transformer +CNN and the ROC curves of each model. The TP, FP, FN, TN of transformer + CNN are 180, 8, 15, 187 respectively. The accuracy of pretrained transformer + CNN is 95.47%, the recall rate is 92.31%, the F1 score is 0.9418, and the AUC is 0.9755. The accuracy of pretrained CNN is 95.81%, the recall rate is 93.85%, the F1 score is 0.9492, and the AUC is 0.9799. It indicates that with the assistance of our model, the probability of missed diagnosis and misdiagnosis could be greatly reduced.

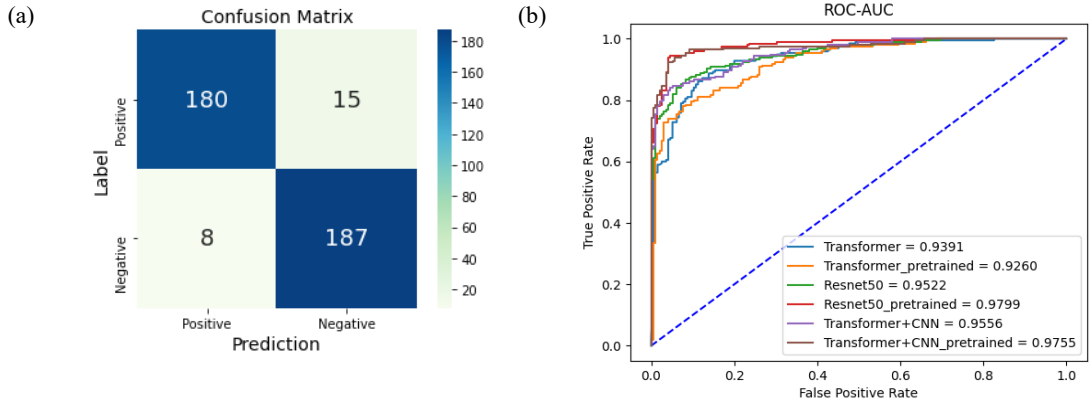

Figure 3. Results of the classification of "benign" and "malignant". (a)Confusion matrix of transformer + CNN. (b)ROC curves and AUC values of all models.

## B. Description Model and Fixed Model

In the validation set, the performance of CNN+SVM feature classification model and VGG network are shown in Table 6 and the specific reference of "class 0" or "class 1" of each feature are marked in Table 2. For example, for internal echo, "class 0" means that the internal echo is homogenous, and "class 1" equals to anechoic.

Table 6 results on validation set

| Feature | SVM(class 0) | SVM(class 1) | VGG(class 0) | VGG(class 1) |
|---|---|---|---|---|
| Internal Echoes | 0.7115 | 0.7278 | 0.5316 | 0.5 |
| Posterior Acoustic | 0.843 | 0.8707 | 0.9206 | 0.9197 |
| Shape | 0.8966 | 0.8571 | 0.7736 | 0.9396 |

Results of accuracy assessments for internal echoes, posterior acoustic features, and mass shape of the five-physician test set are shown in Table 7. At the same time, it should be noted that the evaluation criteria of different doctors could not be unified, and there were some subjective factors for each doctor, but the results showed the validity of the model to some extent.

Table 7 results of accuracy assessments of test set

| Model | NAME | Internal Echoes | Posterior Acoustic | N2 | Shape | N3 | Internal Echoes1 | N1 |
|---|---|---|---|---|---|---|---|---|
| **CNN+SVM** | Doctor1 | 0.4889 | 0.6744 | 43 | 0.9130 | 46 | 0.6923 | 45 |
| | Doctor2 | 0.5122 | 0.8293 | 41 | 0.6585 | 41 | 0.5000 | 41 |
| | Doctor3 | 0.5111 | 0.5000 | 42 | 0.6000 | 45 | 0.7857 | 45 |
| | Doctor4 | 0.5625 | 0.7187 | 32 | 0.6562 | 32 | 0.6129 | 32 |
| | Doctor5 | 0.6154 | 0.4615 | 26 | 0.9615 | 26 | 0.7895 | 26 |
| | **Average** | 0.5289 | **0.6448** | 36.8 | **0.7473** | 38 | **0.6729** | 37.8 |
| **VGG** | Doctor1 | 0.6222 | 0.8139 | 43 | 0.9783 | 46 | 0.6429 | 45 |
| | Doctor2 | 0.4878 | 0.9268 | 41 | 0.7073 | 41 | 0.7561 | 41 |
| | Doctor3 | 0.5000 | 0.8125 | 42 | 0.8750 | 45 | 0.7619 | 45 |
| | Doctor4 | 0.5385 | 0.7692 | 32 | 0.9615 | 32 | 0.8437 | 32 |
| | Doctor5 | 0.6000 | 0.8333 | 26 | 0.9111 | 26 | 0.6087 | 26 |
| | **Average** | 0.5471 | **0.8331** | 36.8 | **0.8833** | 38 | **0.7247** | 37.8 |

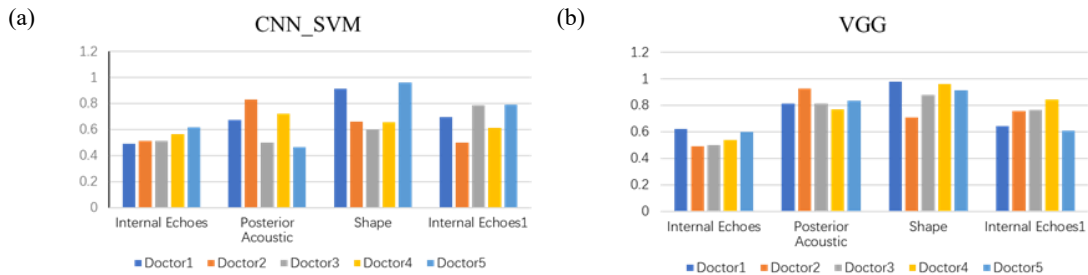

Figure 4. Histogram of test set results by CNN+SVM(a) and VGG(b), respectively. We achieved better results in CNN than SVM.

The histogram of the test set results is shown in Fig. 4. The classification probability of internal echoes is about 0.5 (as shown in the first column "Internal Echoes" of Fig. 4 (a) and (b)), which is similar to random classification. According to clinicians' experience, when the internal echo is anechoic, the mass is usually cystic. Due to the large amount of water in the mass, there is often an enhanced echo behind. Therefore, the feature fusion method was adopted in this paper to classify the internal echoes. The benefit can be observed in the last column "Internal Echoes 1" of histogram Fig.4(a) and (b). We used 1/0 represents yes/no enhancement of the posterior acoustic and 1/0 represents anechoic/homogeneous of the internal echo, thus the results can be divided into four categories, as shown in Table 8. It should be noticed that "00" in the above four results is extremely unlikely to occur, so the remaining conditions were classified into three subcategories.

Table 8 Feature fusion

|  | Homogeneous | Anechoic |
|---|---|---|
| **Enhancement** | 00 | 01 |
| **No posterior features** | 10 | 11 |

According to the above results, a preliminary ultrasound report can be generated with *if-then* rules. For the lesions judged as benign, the orientation of the tumor is parallel to the skin by default, and the margin of the tumor is circumscribed and the boundary is abrupt. These are key features to distinguish the benign from the malignant. And for benign cases, these features are completely consistent, which can be automatically generated in the report without any additional judgment. Examples of breast ultrasound report generated are shown in Fig. 5.

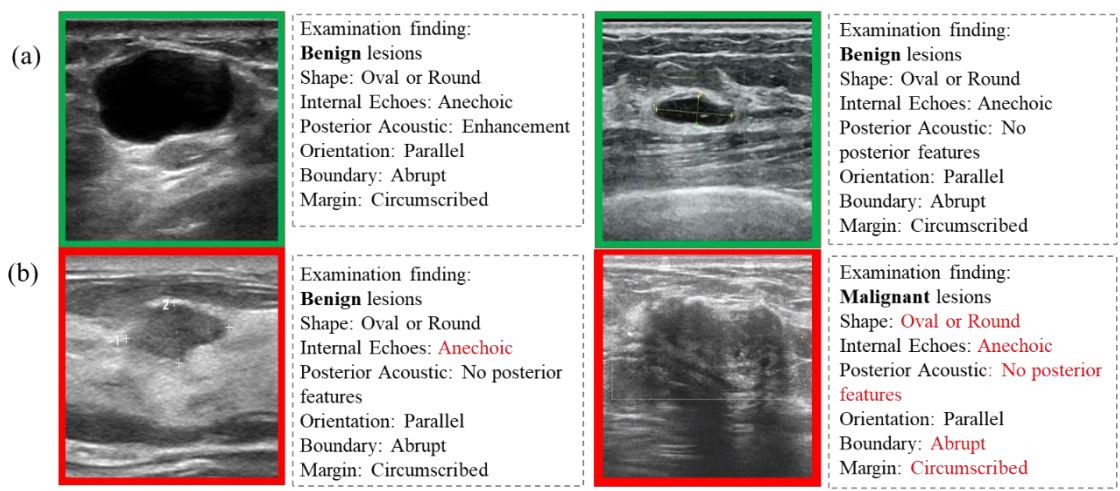

Figure 5. Examples of breast ultrasound report generated. (a) Correct predictions of benign lesions and its features. (b)Red font indicates wrong predictions. Left: The label of internal echoes is homogeneous, but the predicted result is anechoic. Right: This is a case of malignant lesion, the doctor should type the feature descriptors where the red font prompts to generate the final malignant report.

## C. Evaluation index and comprehensive index of all features

The comprehensive accuracy of the model's feature determinations was used as the evaluation index for the improvement of work efficiency, only the cases judged benign by doctors were considered. The statistical result of the six benign features is shown in the last column "Average Of All" of Table 9. It can be seen that for benign cases, compared to nonintelligent structured reports based on fixed templates or with drop-down options, doctors can save more than 90% of the repetitive work that requires manually typing or mouse clicking for selection.

Table 9 Evaluation index and comprehensive index of all features

| Feature | Posterior Acoustic | Shape | Internal Echoes1 | **Average** | Boundary | Orientation | Margin | **Average Of All** |
|---|---|---|---|---|---|---|---|---|
| N | 184 | 190 | 189 | **187.65** | 190 | 190 | 190 | **37.77** |
| Precision | 0.8331 | 0.8833 | 0.7247 | **0.8137** | 1.0 | 1.0 | 1.0 | **0.9074** |

# Ⅴ  Discussion

The innovation of this paper is to effectively extract the image features of benign breast ultrasound cases and automatically generate reporting terms. Different from MRI and CT images, ultrasound images are adopted to generate screening reports. Although ultrasound technology plays an important role in the early screening of breast cancer, it has brought serious burdens to doctors, especially for reports writing. Compared with simple lesion detection or segmentation, our model can greatly reduce nearly 90% of the workload. Personalized report generation is more widely recognized by doctors in breast cancer screening compared to nonintelligent reports based on fixed templates or containing blanks with options to fill in.

During report generation, tasks often aim to produce reports descriptions that are similar to natural language, but this can lead to vastly different descriptions from different doctors, causing distress for patients and trouble in cross hospital diagnostic. On the contrary, structured breast ultrasound report can solve this problem well. This paper focused on feature description according to the BI-RADS standard, which enhances the standardization and consistency of the report.

For structured reports, there is still a lot to improve in the diversity and accuracy of feature description. In addition, our model is not yet able to describe the left/right breast, we can only generate the corresponding report according to the order of the images selected. Distinguishing between left and right breast images remains a challenge for report generation tasks. Furthermore, the existing report generation tasks usually generate descriptions at the picture level, which reduces the possibility of practical application in medical care because one patient may have more than one images at once. To the best of our knowledge, there is no method for patient-level report generation, which is a more challenge direction for future research. In addition, due to the imbalanced data in clinic, it is often difficult to apply experimental methods to practice. Malignant cases, for example, happens rarely in reality, so the risk of misleading prediction associated with the malignant features would significantly increase the amount of work to revise preliminary reports from the malignant to the benign. However, in this paper, the most common features are selected for description to reduce prediction errors caused by model trained with imbalanced data.

Hopefully, because of the efficiency and automation of AI assisted report, we believe that structured and intelligent diagnosis reports can greatly relieve doctors' work pressure and will be widely used in early breast cancer screening.

# Ⅵ  Conclusion

In this paper, an AI-assisted method for doctors to efficiently generate breast ultrasound screening reports was proposed, and reports written by different doctors were used to test the model. Based on AI generation of breast ultrasound diagnosis report, we can not only expedite clinical workflow, but also provide data support for the intellectualization and popularization of structured clinical ultrasound diagnosis reports in the future.

**Ethics Approval and Consent to Participate**
The collection and analysis of subjects' breast ultrasound data in this study were approved by the Ethics Committee of Shenzhen Hospital of Peking University, China and Shenzhen City Baoan District Women's and Children's Hospital, China. (Approval number: ChiCTR2100047368)

**Human or Animal rights**
Not applicable

**Consent for Publication**
Not applicable

**Availability of Data and Materials**
None

**Funding**
We gratefully acknowledge the financial support from Foundation of Shenzhen Science and Technology Planning Project (NO.GJHZ20200731095205015 and NO.JSGG20191129103020960) and the International Cooperation Foundation of Tsinghua Shenzhen International Graduate School (No. HW2021001).

**Conflict of interest**
The Author(s) declare(s) that there is no conflict of interest.

**Acknowledgements**
None